\documentstyle[12pt]{article}
\def\Journal#1#2#3#4{{#1} {\bf #2}, #3 (#4)}
\def\AP{{\em Ann. Phys.}}

\def\JMP{{\em J. of Math. Phys.}}
\def\JPA{{\em J. of Phys.} A}
\def\MAMS{{\em Mem. Am. Math. Soc.}}

\def\PR{{\em Phys. Rev.}}

\def\PRD{{\em Phys. Rev.} D}

\def\PRL{\em Phys. Rev. Lett.}
\def\RMP{{\em Rev. Mod. Phys.}}

\newcommand{\be}{\begin{equation}}
\newcommand{\ee}{\end{equation}}
\newcommand{\bea}{\begin{eqnarray}}
\newcommand{\eea}{\end{eqnarray}}

\newcommand{\hf} {{1\over2}}

\newcommand{\nonu}{\nonumber\\}
\def\ra{\rangle}
\def\la{\langle}
\def\s#1{{\bf#1}}
\def\lb{\lambda\hskip -.2cm -}

\begin{document}
\begin{center}
{\LARGE Path Integral for Relativistic Equations of Motion}\\
\vspace*{1cm}
Pierre Gosselin\footnote{gosselin@math.u-strasbg.fr}
\footnote{Address after Sept. 1997: Dep. of Mathematics,
J. Fourier University, Grenoble, France}$^a$,
Janos Polonyi\footnote{polonyi@fresnel.u-strasbg.fr}$^{bc}$\\
\vspace*{1cm}
$^a$Department of Mathematics, IRMA, Louis Pasteur University\\
4 rue Ren\'e Descartes 67084 Strasbourg, Cedex, France\\
\vspace*{.5cm}
$^b$Laboratory of Theoretical Physics, Louis Pasteur University\\
3 rue de l'Universit\'e 67087 Strasbourg, Cedex, France\\
\vspace*{.5cm}
$^c$Department of Atomic Physics, L. E\"otv\"os University\\
Puskin u. 5-7 1088 Budapest, Hungary\\
\vspace*{2cm}
{\LARGE Abstract}\\
\end{center}
A non-Grassmanian path integral representation is given for
the solution of the Klein-Gordon and the Dirac equations.
The trajectories of the path integral are rendered differentiable
by the relativistic corrections.
The nonrelativistic limit is briefly discussed from
the point of view of the renormalization group.

\section{Introduction}
The path integral representation for the solution of the Schr\"odinger
equation is not only a powerful computational method but provides a
framework to the understanding of the propagation in nonrelativistic
Quantum Mechanics. In a surprising manner the known extensions of
this formalism for spin half or relativistic particles are not
too satisfactory.

Due to the spin-statistics theorem \cite{finkrub}
the second quantized path integral for fermions is usually given in
terms of anticommuting, formal Grassmann variables. The first
quantized path integral for the Dirac equation can be obtained
by means of the Grassmann variables in a manner analogous to
the case of the Brownian motion \cite{gavsch}. In order to
apply local nonperturbative methods to the path integral one needs a
less formal basis in terms of ordinary complex numbers.
A Brownian motion on a rectangular space-time lattice with a
Poisson distributed helicity flip was used to construct a c-number path 
summation in ref. \cite{jacsch} which satisfies the Dirac equation. 
The underlying random walk was eucidated in \cite{gjks} but without
path integration. The application of this formalism is not clear 
in the more realistic cases.

The spin zero relativistic equation of motion is of second order in the
time derivatives which creates problems in the generalization of
the nonrelativistic path integral. Only
Schwinger's proper time formalism \cite{schw} is known to
tackle this problem in the second quantized formalism.

We outline in this paper a modified version of the path integration
which reproduces the solutions of the Dirac and the Klein-Gordon
equations. It is similar to the standard path integral known from
the Schr\"odinger equation except the lagrangian is matrix valued. 
Though this feature makes it less useful and attractive than Feynman's
original path integral however the ordinary integration over the 
coordinates may in the future lead to an eventual discovery of a more attractive
path integral formalism for fermions. The characteristic feature of the construction is
that the cutoff of the path integral, $\Delta t$, remains
explicitly present in the continuum limit, $\Delta t\to0$.
In this respect our construction is formally similar to the
second quantized Quantum Field Theory models which require
the introduction of a cutoff due to their ultraviolet divergences
even if this cutoff, as in our case, can be pushed beyond
any finite limit without modifying the physical content of
the theories.

The main achievent of the path integral formalism is that the
particle is perfectly localised along the trajectories during
the propagation. The relevance of this formalism is naturally limited up to the
energies of the transition between the relativistic and the nonrelativistic
regime. At distances below the Compton wave length the localization
becomes wague what can be seen physically in the creation of the
particle-anti particle pairs or mathematically in the smearing of
the relativistically covariant coordinate operator \cite{wigner}
and second quantization is needed. Nevertheless our
representation of the solutions of the relativistic equations of motion
may provide a conceptional insight into the transition between the
relativistic and the nonrelativistic region, the origin of the spin
in the first quantized description and might be useful as a
framework for nonperturbative approximations as a starting
point to the relativistic valence approximation.
We shall consider free particle for simplicity.

The organization of the paper is the following. We recapitulate some
properties of the nonrelativistic path integral in Section 2. Section 3
contains
the derivation of the path integral representation for the
solution of the massless Dirac equation. The massive case is
dealt with in Section 4. The finite time propagator
is investigated in Section 5 by means of the
Foldy-Wouthuysen transformation. The path integral for the
solution of the Klein-Gordon equation is presented in Section 6.
Section 7 is a brief qualitative digression to comment our results
from the point of view of the renormalization group
and to verify the nonrelativistic limit.
Finally Section 8 is reserved for the summary.

\section{Nonrelativistic path integral}
In order to motivate our construction of the relativistic path integration 
we start with a hand-waving introduction of the nonrelativistic path integral 
for a free particle in flat space what is based on the assumtion that the action is a 
quadratic functional. The propagator,
\be
\la\s x|e^{{it\hbar\over2m}\partial^2}|\s y\ra=G_0(\s x,\s y;t)
=e^{{i\over\hbar}S(\s x,\s y;t)},
\ee
is always self reproducible, i.e. a fixed point of the convolution,
\be
e^{{i\over\hbar}S(\s x,\s y;t_1+t_2)}=\int d\s z
e^{{i\over\hbar}S(\s x,\s z;t_1)}e^{{i\over\hbar}S(\s z,\s
y;t_2)}.\label{gblock}
\ee
The asumption is that this is achieved in the simplest manner, i.e. by choosing
quadratic expression for $S(\s x,\s y;t)$ in the coordinates. 
The form of $S(\s x,\s y;t)$ is then further constrained by the kinematical symmetries, 
i.e. invariance under the translations in space and time, the three-rotations and the space
and time inversion. 

The translation symmetry restricts the functional form to
\be
S(\s x,\s y;t)=\hf(x-y)^jA_{jk}(x-y)^k+(x-y)^jB_j.\label{rectrf}
\ee
Rotational symmetry imposes $A_{jk}=A\delta_{j,k}$ and $\s B=0$ for
spin zero particles described by single component wave functions. In the case
of spin the rotational symmetry allows $\s B=B\s S$, where $\s S$ is the
spin operator but the space inversion symmetry requires $B=0$. The dimension
of the action is $[S(\s x,\s y;t)]=ML^2T^{-1}$ so $[A]=MT^{-1}$. Due to the
translation invariance in time there is no other time dimensional parameter
in the action than $t$ so $A=m/t$. The quantity $m$ has mass dimension and
is real due to the time inversion symmetry.

In the presence of an external potential the action is nonquadratic and
the symmetry properties are not enough
to determine the propagator. Direct computation yields in the short time limit
the well known expression, \cite{fh},
\bea
G(\s x,\s y;\Delta t)&=&\biggl({m\over2\pi i\hbar\Delta t}\biggr)^{3\over2}\\
&&\times\exp\left({im(\s x-\s y)^2\over2\hbar\Delta t}
-{i\over\hbar}\Delta tV\left({\s x+\s
y\over2}\right)\right)\nonumber\label{kernel}
\eea
for the hamiltonian $H={\s p^2\over2m}+V(\s x)$. The simplest proof
of this result is to verify the time evolution of the wave function,
\be
\psi(\s x,t+\Delta t)=\int d\s yG(\s x,\s y;\Delta t)\psi(\s y,t).
\label{prop}
\ee
In fact, this reproduces the Schr\"odinger equation as $\Delta t\to0$,
\bea
\psi(\s x,t+\Delta t)&=&\biggl[1-{i\over\hbar}\Delta t\biggl(
-{\hbar^2\over2m}\partial^2+V(\s x)\\
&&+{i\hbar\Delta t\over2m}\partial V(\s x)\cdot\partial
+\cdots-{i\hbar^3\Delta
t\over8m^2}(\partial^2)^2+\cdots\biggr)\biggr]\psi(\s x,t).
\nonumber
\eea
The first set of dots stands for the contributions which are higher order
in the Taylor
expansion of the potential $V(\s x)$. The second set of dots represents the
higher
orders in the expansion of the wave function. Let us introduce
\be
\lb_B={\hbar\over|\s p|},
\ee
the de Broglie wave length divided by $2\pi$, the
characteristic scale of the the wave function,
The small parameter in the first set of corrections is
\be
{{\Delta t\hbar|\partial V|\over\lb_B m}\over{\s p^2\over2m}}
={{\Delta v\hbar\over\lb_B}\over{\s p^2\over2m}}
={\Delta\left({\s p^2\over2m}\right)\over{\s p^2\over2m}},
\ee
where we used the classical relations
\be
m\Delta v=\Delta p=\partial V\Delta t.
\ee
These corrections are small if the kinetic energy does not change too much
during the time $\Delta t$ of the propagation. The second kind of corrections
are organized according to the small parameter
\be
{{\hbar^3\Delta t\over m^2\lb_B^4}\over{\s p^2\over m}}
={|\s v|\Delta t\over4\pi^3\lb_B},\label{freec}
\ee
indicating that the particle should drift by a small amount compared to its
characteristic length during the time $\Delta t$. These kind of corrections
can be reexpontentiated and pose no problem for the free
nonrelativistic propagation.
In our study of the free particle we shall keep track of such
kind of corrections only.

The expression (\ref{kernel}) is not
unique, the method of the renormalization group can be used to
identify the family of lagrangians which belong to the universality
class of the Schr\"odinger equation \cite{rgqm}.

\section{Massless fermions}
The covariant equation of motion for the wave function $\phi$ of the
first quantized massless spin half particle (right handed anti-neutrino) is
\be
(p^0-c\s
p\cdot\sigma)\phi=(i\hbar\partial_0+ic\hbar\partial\cdot\sigma)\phi=0.
\label{neut}
\ee

Since the space inversion
\be
{\cal P}\s x\longrightarrow-\s x,
\ee
does not commute with the Lorentz boosts
$\cal P$ can not be represented trivially in the irreducible representation
$\phi$ of the special Poincare group. After space inversion one finds
\be
(p^0+c\s
p\cdot\sigma)\chi=(i\hbar\partial_0-ic\hbar\partial\cdot\sigma)\chi=0,
\label{antin}
\ee
with
\footnote{The factor $i$ included here is a convention as long as
there is no neutral spin half particle.}
\bea
{\cal P}\phi&=&i\chi,\nonu
{\cal P}\chi&=&i\phi.
\eea
The equation (\ref{antin}) is covariant only if $\chi$ (left handed neutrino)
transforms as a complex conjugate spinor under the Lorentz transformations.

Our requirements about the propagator for finite time
are: (i) Self reproducibility under the convolution, (\ref{gblock}), achieved by 
a quadratic action, (ii) Covariance with respect to the special Poincare group
and (iii) the restriction of the propagation along the light cone even for
the quantum fluctuations, i.e. independently of $\hbar$ up to the distance
scale of the cutoff, $c\Delta t$. As in the nonrelativistic case, (i)
and the invariance under translations and rotations in space leads to the form
(\ref{rectrf}). Since the space inversion is not expected to be a symmetry,
$B_j=B\sigma_j$. The strong kinematical constraint on the propagation which is
imposed even on the level of the quantum fluctuations requires that the
infinitesimal time propagator should be independent of
$\hbar$. Since there is no parameter with mass
dimension
the only choice is
\be
G(\s x,\s y;\Delta t)={\cal N}^{-1}(\Delta t)
\exp\left(i{A\over2}\left({\s x-\s y\over c\Delta t}\right)^2
-iB{\s x-\s y\over c\Delta t}\cdot\sigma\right),
\ee
where $\cal N$ is fixed by the normalization, $A$ and $B$ are dimensionless
numbers. By analogy with the nonrelativstic case the $-i\hbar$ times the
exponent in this expression might be called the action. We set $B=1$ in order
reproduce the propagation where $c$ is the speed of light. One finds for the
time evolution of the amplitude,
\bea
\phi(\s x,t+\Delta t)&=&\int d\s yG(\s x,\s y;\Delta t)\phi(\s y,t)\\
&=&{\cal N}^{-1}e^{i{3\over2}}(2\pi ic^2\Delta t^2)^{3\over2}\nonu
&&\times\biggl(1-{c\Delta t\over A}\sigma\cdot\partial
+{3c^2\Delta
t^2\over2A}\left(A^{-1}+i\right)\partial^2+\cdots\biggr)
\phi(\s x,t),\nonumber\label{neqd}
\eea
what agrees with the equation of motion as $\Delta t\to0$ for $A=1$ as long
as the characteristic length, $\lb_B$, of the amplitude $\phi$ is long
enough c.f. (\ref{freec}),
\be
c\Delta t<<\lb_B.\label{linc}
\ee
The normalization yields
\be
{\cal N}=e^{i{3\over2}}(2\pi ic^2\Delta t^2)^{3\over2}.
\ee
Thus the path integral
\bea
G(\s x,\s y;t)&=&\prod_j\left\{{\cal N}^{-1}\int d\s x_j\exp
\left[{i\over2}\left({\s x-\s y\over c\Delta t}\right)^2
-i{\s x-\s y\over c\Delta t}\cdot\sigma\right]\right\}\nonu
&=&\int D[\s x(t)]T\exp
\left\{i\int dt\left[\hf\left({d\s x\over dt}\right)^2{1\over c^2}{1\over dt}
-{1\over c}{d\s x\over dt}\cdot\sigma\right]\right\}\label{pathiml}
\eea
over the trajectories for time $t$ with the end points $\s x$ and $\s y$ 
reproduces the propagator for finite time.
The time ordering $T$ in the second line takes care of the noncommutation
of the spin mixing at different time slices.

Notice the
explicit linear divergence, ${1\over dt}$, in the action. The role of the
divergence is to suppress the term $\partial^2$ in the equation of motion.
It is impossible to maintain the linear equation of motion without keeping
the cutoff, $\Delta t$, large but finite. This is reminiscent of Quantum Field
Theory where the bare lagrangian contains the divergent coupling constants.
Eq. (\ref{neqd}) assures that the hamiltonian corresponding this
diverging bare lagrangian is actually renormalizable and finite,
\be
H=-ic\hbar\partial\cdot\sigma.\label{hamiltml}
\ee

The path integral (\ref{pathiml}) can easily be computed by the help of the
Fourier transformation
\bea
\tilde G(\s p,\Delta t)&=&\int d\s ze^{-{i\over\hbar}\s x\s p}
G(\s x,0;\Delta t)\nonu
&=&\exp\left[-i\left(\hf\left({\s pc\Delta t\over\hbar}\right)^2
-{\s pc\Delta t\over\hbar}\cdot\sigma\right)\right],
\eea
since the convolution (\ref{gblock}) becomes multiplication in the momentum
space,
\bea
G(\s x,\s y;t_1+t_2)&=&\int d\s z{d\s p_1d\s p_2\over(2\pi\hbar)^6}
e^{{i\over\hbar}(\s x-\s z)\s p_1+{i\over\hbar}(\s z-\s y)\s p_2}
\tilde G(\s p_1,t_1)\tilde G(\s p_2,t_2)\nonu
&=&\int{d\s p\over(2\pi\hbar)^3}e^{{i\over\hbar}(\s x-\s y)\s p}
\tilde G(\s p,t_1)\tilde G(\s p,t_2).
\eea
The Fourier transformed propagator for time $t$ is
\be
\tilde G(\s p,t)=\exp\left[-it\left(\hf\left({\s pc\over\hbar}\right)^2
\Delta t-{\s pc\over\hbar}\cdot\sigma\right)\right],
\ee
what yields
\be
G(\s x,\s y;t)={\cal N}^{-1}
\exp\left[{t\over\Delta t}\left({i\over2}\left({\s x-\s y\over ct}\right)^2
-i{\s x-\s y\over ct}\cdot\sigma\right)\right].\label{kernelft}
\ee
The energy of the states described by the plane wave with
momentum $\s p$ is
\be
E(\s p)=\pm|\s p|c\left(1+O\left({\Delta tc\over\lb_B}\right)\right).
\ee
Note that the commutativity
\be
[\tilde G(\s p,t),\s p\cdot\sigma]=0
\ee
indicates that the helicity is conserved during the propagation.

It is illuminating to rewrite the logarithm of the propagator for finite
time as
\be
S(\s x,\s y;t)={\tilde m(\s x-\s y)^2\over2t}-{1\over c}(\s x-\s y)\cdot\s A,
\label{nrela}
\ee
where $\tilde mc^2=\hbar\omega/2\pi$, $\omega=2\pi/\Delta t$,
$\s A=\s A^a\sigma^a/2$, and $A^a_j=\delta^a_j\tilde mc^2$.
This shows a formal similarity with the propagation of a
particle with mass $m$ in the presence of an SU(2) background field,
$\s A$, whose magnetic field strength is
\be
B^a_j=\delta^a_j\tilde m^2c^4
\ee
in nonrelativistic Quantum Mechanics. 

In deriving the neutrino equation we had to neglect the term $O(\partial^2)$
what required that the cutoff, $c\Delta t$, be smaller than
the characteristic length scale of the amplitude, $\lb_B$. So long
as the cutoff is kept small but finite the path integral contains 
some undesired, noncovariant contributions at the distance scales
below $c\Delta t$. This is obvious from (\ref{kernelft}) where the exponent
is slowly varying at such distance scales thus the propagation in not 
constrained on the light cone. For longer distances the phase of the exponent is
large and its stationary point with respect $\s x$ yields the "equation of motion"
\be
{\s x-\s y\over ct}\phi(\s y,0)=\sigma\phi(\s y,0).
\ee
This guarantees that the particle stays on the light cone because the
eigenvaules of the Pauli matrices is $\pm1$.

In order to understand the role of the cutoff better we follow the spread of 
a wave packet with the Fourier transform
\be
\tilde\phi(\s p)=(2\pi\Delta p^2)^{-{3\over2}}
\exp\left(-{\s p^2\over2\Delta p^2}\right)\phi_0,\label{initc}
\ee
at $t=0$ where $\phi_0^\dagger\phi_0=1$. The wave function at time $t$ is then
\bea
\phi(\s x,t)&=&(2\pi\sigma^2)^{-{3\over2}}\int{d\s p\over(2\pi\hbar)^3}\nonu
&&\exp\left[{i\over\hbar}\s p\cdot\s x
-it\left(\hf\left({\s pc\over\hbar}\right)^2
\Delta t+{\s pc\over\hbar}\cdot\sigma\right)-{\s p^2\over2\Delta p^2}
\right]\phi_0\nonu
&=&{\cal N}^{-\hf}\exp\left(-\hf A(\s x^2-2tc\s x\cdot\sigma)\right)\phi_0,
\eea
where ${\cal N}$ is a time dependent constant and
\be
A={1+it\Delta tc^2{\Delta p^2\over\hbar^2}\over{\hbar^2\over\Delta p^2}
+(t\Delta tc^2)^2{\Delta p^2\over\hbar^2}}.
\ee
The absolute magnitude of the wave function is
\be
\phi^\dagger(\s x,t)\phi(\s xt)
={\cal N}^{-1}(t)\exp\left(-{\s x^2\over2\Delta x^2(t)}\right),
\ee
with
\bea
\Delta x^2(t)&=&\Delta x^2(0)+(tc)^2{(\Delta tc)^2\over\Delta x^2(0)},\nonu
\Delta x^2(0)&=&\lb_B^2={\hbar^2\over\Delta p^2}.\label{spread}
\eea
By comparing this result with the spread of the nonrelativistic
wave packet we find that that $\Delta x^2(t)$ corresponds to the usual 
nonrelativistic
expression obtained from (\ref{nrela}) and is independent of $\hbar$
when expressed in terms of $\Delta x^2(0)$. Furthermore the explicit
apparence of the cutoff in the finite time propagator, (\ref{kernelft}),
induces acausal propagation by the gradual loss of coherence between 
the high momentum modes when the intial wave function varies considerably 
within the cutoff. This spread is suppressed and the propagation
becomes causal for $\Delta tc<<\Delta x$.

The lesson to be learned from (\ref{spread}) is that the wave function is 
constrained by causality only as $\Delta t\to0$. The acausal propagation 
is unaccessible for observables far from the cutoff and drops
out from the "renormalized" theory, where $\Delta t=0$. How does this
propagation look like? It turns out to be similar to the nonrelativistic case.
In fact, at short distances, i.e. high momentum the $O(\partial^2)$ part
of the time evolution is the dominant one and we find a 
nonrelativistic propagation. According to (\ref{nrela}) 
the mass parameter, $\tilde m$, of this propagation
is such that its Compton wave length is just the cutoff, $c\Delta t$.
Another role the short distance modes play is
that the phase of the wave function which diverges for a perfectly localized
state, $\lb_B\to0$, becomes finite when (\ref{linc}) holds
due to the interference between points with distance $c\Delta t$
in the original wave packet. The phase of the amplitude
for wave packet with finite extent is cutoff independent in this manner.

\section{Massive fermions}
The mass of a spin half particle is the parameter which controls the
mixing between a particle and its space inverted state,
\bea
(p^0-c\s p\cdot\sigma)\phi&=&mc^2\chi,\nonu
(p^0+c\s p\cdot\sigma)\chi&=&mc^2\phi.
\eea
This mixing can be reproduced by multiplying the massless kernel
for the bispinor $\psi=(\phi,\chi)$ by
\be
\exp\left(-{mc^2\Delta t\over\hbar}{\cal P}\right).
\ee
The kernel for a massive fermion is then
\bea
G(\s x,\s y;\Delta t)&=&e^{-{3\over2}i}(2\pi ic^2\Delta t^2)^{-{3\over2}}\\
&&\times\exp\left[i\left(\hf\left({\s x-\s y\over c\Delta t}\right)^2
-\alpha\cdot\left({\s x-\s y\over c\Delta t}\right)\right)\right]\nonu
&&\exp\left(-{i\over\hbar}\beta mc^2\Delta t\right),\nonumber\label{massivk}
\eea
where
\bea
\beta&=&\gamma^0=\pmatrix{0&1\cr1&0},\nonu
\alpha^j&=&\gamma^0\gamma^j=\pmatrix{\sigma^j&0\cr0&-\sigma^j}.
\eea
By substituting this expression into (\ref{prop}) we obtain
\be
\psi(\s x,t+\Delta t)=\left(1-{i\over\hbar}\beta mc^2\Delta t
-\Delta tc\alpha\cdot\partial-{1\over8}c^2\Delta
t^2\partial^2+\cdots\right)\psi(\s x,t),\label{infidr}
\ee
what goes over the Dirac equation,
\be
i{\hbar\over c}\partial_0\Psi=\left(-i\hbar\alpha_j\partial_j+\beta
mc\right)\psi,
\ee
as long as (\ref{linc}) holds to suppresses the acausal $O(\partial^2)$
term in (\ref{infidr}) and 
\be
c\Delta t<<\lambda_C={\hbar\over mc}
\ee
in order to perform the linearization in the space inversion. The
latter condition is needed to arrive at a finite difference equation
in time what amounts to the construction of the Poissonian stochastic
process in \cite{gjks}.

\section{Finite Time Propagator}
The propagator for finite time is the initial and the final point
dependent form of the path integral
\bea
G(\s x,\s y;t)&=&\prod_j\left\{\int d\s x_j{\cal N}^{-1}(\Delta t)\exp
\left[{i\over2}\left({\s x-\s y\over c\Delta t}\right)^2
-{\s x-\s y\over c\Delta t}\cdot\alpha\right]\right\}\nonu
&&\times\exp\left(-{i\over\hbar}\beta mc^2\Delta t\right),\nonu
&=&\int D[\s x(t)]T
\exp\left\{i\int dt\left[
\hf\left({d\s x\over dt}\right)^2{1\over c^2}{1\over dt}
-{1\over c}{d\s x\over dt}\cdot\alpha\right]\right\}\nonu
&&\times\exp\left(-{i\over\hbar}\int dt\beta mc^2\right),\label{pathim}
\eea
and it will be obtained by the help of (\ref{gblock}),
\be
G(\s x,\s y;t_1+t_2)=\int d\s zG(\s x,\s z;t_1)G(\s z,\s y;t_2).
\ee
We shall work in Fourier space where the convolution is the multiplication,
\be
\tilde G(\s p,t_1+t_2)=\tilde G(\s p,t_1)\tilde G(\s p,t_2).\label{fblock}
\ee

The finite time propagator can not be given in
terms of the exponential of a simple quadratic expression due
the nonlinear time dependent mixing of the chiral spinors. But a partial
simplification is gained by the Foldy-Wouthuysen transformation given by
the similarity transformation
\be
U_{FW}=\exp\left(-{i\over2mc}\beta\alpha\cdot\s p\right)
\left(1+O\left({c\Delta t\over\lb_B}\right)\right)
\ee
which decouples the large and the small components of the standard
representation.

One can perform the Fourier transformation with the result
\be
\tilde G(\s p,\Delta t)
=\exp\left[-i\left(\hf\left({|\s p|c\Delta t\over\hbar}\right)^2
-{c\Delta t\over\hbar}\alpha\cdot\s p\right)\right]
\exp\left(-{i\over\hbar}\beta mc^2\Delta t\right).
\ee
The convolution (\ref{fblock}) suggests that the 
relativistic-nonrelativistic crossover will be at $c\Delta t\approx\lambda_C$
where the noncommutativity due to the mass term becomes important.
By the help of the generalized Euler relation,
\be
e^{i\s v\cdot\alpha}=\cos|\s v|+i\alpha\cdot{\s v\over|\s v|}\sin|\s v|,
\ee
we can write
\bea
\tilde G(\s p,\Delta t)&=&
\exp\left(-i\hf\left({|\s p|c\Delta t\over\hbar}\right)^2\right)\nonu
&&\times\left[\cos\left({|\s p|c\Delta t\over\hbar}\right)
+i\alpha\cdot{\s p\over|\s p|}\sin\left({|\s p|c\Delta
t\over\hbar}\right)\right]\nonu
&&\times\exp\left(-{i\over\hbar}\beta mc^2\Delta t\right).
\eea

We now introduce the three rotation $R(\s p)$ which rotates the
momentum into  the $z$-direction, $R(\s p)\s p=\hat z$ and thus diagonalize
the
Fourier transformed infinitesimal time propagator. In the spin half
representation
\be
R(\s p)\longrightarrow R_{1/2}(\s p)=e^{-{i\over2}\phi\s n\cdot\sigma},
\ee
and one finds
\be
R_{1/2}(\s p)\s p\cdot\sigma R_{1/2}^{-1}(\s p)
=|\s p|\pmatrix{1&0\cr0&-1\cr}.
\ee
This allows us to write
\be
\tilde G(\s p,\Delta t)=e^{-{i\over2}\left({c\Delta t\over\lb_B}\right)^2}
\pmatrix{R^{-1}_{1/2}&0\cr 0&R^{-1}_{1/2}\cr}{\cal G}
\pmatrix{R_{1/2}&0\cr0&R_{1/2}\cr}
\ee
with
\be
{\cal G}=\pmatrix{z\cos\left({c\Delta t\over\lb_B}\right)&0&
+iz^*\sin\left({c\Delta t\over\lb_B}\right)&0\cr
0&z\cos\left({c\Delta t\over\lb_B}\right)&0&
-iz^*\sin\left({c\Delta t\over\lb_B}\right)\cr
+iz\sin\left({c\Delta t\over\lb_B}\right)&0&
z^*\cos\left({c\Delta t\over\lb_B}\right)&0\cr
0&-iz\sin\left({c\Delta t\over\lb_B}\right)
&0&z^*\cos\left({c\Delta t\over\lb_B}\right)\cr},
\ee
expressed in the standard representation and
\be
z=e^{-i{c\Delta t\over\lambda_C}}.
\ee

The eigenvalues $\lambda$ of $\cal G$ satisfy the characteristic equation
\be
{\rm Det}\left[G(p,\Delta t)-\lambda{\rm Id}\right]
=\left[1+\lambda^2-2\cos\left({mc^2\Delta t\over\hbar}\right)
\cos\left({|\s p| c\Delta t\over\hbar}\right)\lambda\right]^2=0.
\ee
There are two double eigenvalues,
\bea
\lambda_{\pm}&=&\cos\left({mc^2\Delta t\over\hbar}\right)
\cos\left({|\s p|c\Delta t\over\hbar}\right)
\pm i\sqrt{1-\left[\cos\left({mc^2\Delta t\over\hbar}\right)
\cos\left({|\s p|c\Delta t\over\hbar}\right)\right]^2}\nonu
&=&\cos\left({mc^2\Delta t\over\hbar}\right)
\cos\left({|\s p|c\Delta t\over\hbar}\right)\nonu
&&\pm i\sqrt{\sin^2\left({|\s p|c\Delta t\over\hbar}\right)
+\cos^2\left({|\s p|c\Delta t\over\hbar}\right)
\sin^2\left({mc^2\Delta t\over\hbar}\right)}
\eea
with unit absolute magnitude.
The matrix $\tilde G(\s p,\Delta t)$ can be transformed into the diagonal
matrix
\be
\tilde G_d(\s p,\Delta t)=\exp
\left(-i\Delta t\beta\theta-{i\over2}\left({|\s p|c\Delta
t\over\hbar}\right)^2\right),
\ee
with $e^{i\Delta t\theta}=\lambda_+$ by means of the Foldy-Wouthuysen
transformation
and $R(\s p)$. The Fourier transform of the $N$-th
power of the infinitesimal time propagator is
\bea
\tilde G^N(\s p,\Delta t)&=&\tilde G(\s p,t)\nonu
&=&\exp\left(-{i\s p^2c^2t\Delta t\over2\hbar^2}\right)\nonu
&&\times U_{FW}\pmatrix{R^{-1}_{1/2}&0\cr 0&R^{-1}_{1/2}\cr}e^{-it\beta\theta}
\pmatrix{R_{1/2}&0\cr0&R_{1/2}\cr}U_{FW}^\dagger.\label{mpropf}
\eea

To compute $\theta$ we write $\lambda_+$ as
\bea
\lambda_+&=&\cos\left({mc^2\Delta t\over\hbar}\right)
\cos\left({|\s p|c\Delta t\over\hbar}\right)\nonu
&&\times\left\{1+\tan\left({mc^2\Delta t\over\hbar}\right)
i\sqrt{1+\left[{\tan\left({|\s p|c\Delta t\over\hbar}\right)\over
\sin\left({mc^2\Delta t\over\hbar}\right)}\right]^2}\right\},
\eea
what yields
\bea
\tan\Delta t\theta&=&\tan\left({mc^2\Delta t\over\hbar}\right)
\sqrt{1+\left[{\tan\left({|\s p|c\Delta t\over\hbar}\right)\over
\sin\left({mc^2\Delta t\over\hbar}\right)}\right]^2}\nonu
&=&\tan\left({c\Delta t\over\lambda_C}\right)
\sqrt{1+\left[{\tan\left({c\Delta t\over\lb_B}\right)\over
\sin\left({c\Delta t\over\lambda_C}\right)}\right]^2}.\nonu
\eea
Assuming that the cutoff is chosen to be sufficiently deeply in the
relativistic regime, $c\Delta t<<\lambda_C$, and far from the momentum,
$c\Delta t<<\lb_B$, one obtains
\be
\theta={mc^2\Delta t\over\hbar}\sqrt{1+\left({\s p\over mc}\right)^2}.
\ee
The $N$-th iteration leads then to
\bea
\tilde K_d(\s p,t)&=&\exp\left(-{i\over2N}\left({\s
pct\over\hbar}\right)^2\right)
\exp(-iN\beta\theta)\nonu
&=&\exp\left(-{i\over2N}\left({\s pct\over\hbar}\right)^2\right)\nonu
&&\times\exp\left[-i\beta{mc^2t\over\hbar}\sqrt{1+\left({\s p\over
mc}\right)^2}
+O\left(\sqrt{\Delta t\over t}\right)\right],\label{diagpr}
\eea
where $t=N\Delta t$.

The final expression for the propagator is then
\bea
G(\s x,\s y;t)&=&\int{d\s p\over(2\pi\hbar)^3}
U_{FW}\pmatrix{R^{-1}_{1/2}&0\cr 0&R^{-1}_{1/2}\cr}\nonu
&&\times\exp\left[{i\over\hbar}\s p(\s x-\s y)
-{i\over\hbar}\beta tmc^2\sqrt{1+\left({\s p\over mc}\right)^2}\right]\nonu
&&\times\pmatrix{R_{1/2}&0\cr0&R_{1/2}\cr}U_{FW}^\dagger,\label{print}
\eea
what reproduces the usual relativistic dispersion relation.

\section{Scalar Particle}
The solutions of the equations with second order derivatives in the time have
no direct path integral representation. But it is straightforward to bring the
\be
\left(\Box+{1\over\lambda_C^2}\right)\phi(x)=0,
\ee
Klein-Gordon equation into a first order equation of motion,
\be
{1\over c}\partial_0\Phi_0=\partial_j\Phi_j-m\Phi_d,\label{lkg}
\ee
$j=1,\cdots,d-1,$ for the $d+1$ component field
\be
\Phi=\pmatrix{\Phi_0\cr\Phi_j\cr\Phi_d}
=\pmatrix{{1\over c}\partial_0\phi\cr\partial_j\phi\cr{1\over\lambda_C}\phi}.
\ee
One can verify that the relations
\bea
{1\over c}\partial_0\Phi_d&=&{1\over\lambda_C}\Phi_0\nonu
\partial_j\Phi_d&=&{1\over\lambda_C}\Phi_j,\label{kginic}
\eea
are preserved by (\ref{lkg}).
In order to bring the linearized equation in the form similar to
the Dirac equation we write
\be
i{\hbar\over c}\partial_0\Phi(x)=(-i\hbar\alpha_j\partial_j+\beta
mc)\Phi(x),\label{linkl}
\ee
with the help of the hermitean matrices
\be
\alpha_j=\pmatrix{0&-e_j&0\cr-e_j&0&0\cr0&0&0}~~~~
\beta=\pmatrix{0&0&-i\cr0&0&0\cr i&0&0},
\ee
where $e_j$ is the unit vector in the direction $j$.
An important property of the matrices $\alpha_i$ what holds for any
vector $\bf u$ is
\be
(\s \alpha\cdot\s u)^2=M_u\s u^2,
\ee
with
\be
M_u=\pmatrix{1&0&0\cr0&P_u&0\cr0&0&0}
\ee
where $P_u$ is the projection on $\s u$. $M_u$ satisfies $M_u^2=M_u$.

Analogously with the Dirac equation we introduce the following kernel
\bea
G(\s x,\s y;\Delta t)&=&e^{-{3\over2}iM_u}(2\pi ic^2\Delta t^2)^{-{3\over2}}\\
&&\times\exp\left[i\left(\hf\left({\s x-\s y\over c\Delta t}\right)^2
-\alpha\cdot\left({\s x-\s y\over c\Delta t}\right)\right)\right]\nonu
&&\exp\left(-{i\over\hbar}\beta mc^2\Delta t\right),\nonumber
\eea
In order to obtain (\ref{linkl}) we write in first order in $\Delta t$
\bea
\Phi(\s x,t+\Delta t)&=&\biggl\{1-{i\over\hbar}\beta mc^2\Delta t
-ie^{-{3\over2}iM_u}(2\pi ic^2\Delta t^2)^{-{3\over2}}\nonu
&&\times\int dye^{{i\over2}\left({\s x-\s y\over c\Delta t}\right)^2}
\biggl[M_u{\s\alpha}\cdot\left({\s x-\s y\over|\s x-\s y| }\right)
\sin\left({|\s x-\s y|\over c\Delta t}\right)\nonu
&&+(1-M_u){\bf\alpha}\cdot\left({\s x-\s y\over c\Delta t }\right)\biggr]
\cdot\partial+\cdots\biggr\}\Phi(\s y,t).
\eea
The last equation turns out to be
\bea
\Phi(\s x,t+\Delta t)&=&\biggl(1-{i\over\hbar}\beta mc^2\Delta t
-M_uc\Delta t\s\alpha\cdot\partial\nonu
&&-(1-M_u)e^{-i{3\over2}M_u}c\Delta
t\s\alpha\cdot\partial+\cdots\biggr)\Phi(\s x,t),
\eea
leading to
\be
{\partial\over\partial t}\Phi(\s x,t)=\left(-{i\over\hbar}\beta
mc^2-c\s\alpha\cdot\partial
+\cdots\right)\Phi(\s x,t).
\ee

\section{Scaling laws}
Certain aspects of our construction can be better understood by
means of ideas borrowed from the renormalization group.
We start with the remark that the dimensional analysis
alone is enough to understand the singular properties of the propagation in
Quantum Mechanics. In fact, the action in (\ref{kernel}) contains the term
$(\s x-\s y)^2/t$ so the phase velocity is diverging at short time,
\be
\left({\Delta x\over t}\right)^2\approx{\hbar\over m t},\label{nrelsc}
\ee
reflecting the fast spread of the wave function for well localized
states. This result is not so surprising by recalling that there is
no internal velocity parameter for a free nonrelativistic particle.
In fact, the quantities with the dimension of the velocity have to
be made up by the help of the observational time scale, $t$, which leads to
the divergence.

We may look at Quantum Mechanics as  a Quantum Field Theory in
0+1 dimensions. The ultraviolet divergences make
the introduction of a cutoff necessary in Quantum Field Theory where
a model is renormalizable if this artificial parameter, the cutoff, can be
pushed beyond any limit and can thus be removed. Nevertheless, the precise
mathematical construction of the model requires the introduction of the
cutoff with arbitrary high but finite value. The cutoff can completely
be eliminated in the nonrelativistic systems without
velocity dependent interactions, \cite{rgqm}, and the path integral
can formally be written in the continuum as
\be
\int D[\s x]e^{{i\over\hbar}\int dtL[\s x]},
\ee
where $L[\s x]$ is the well defined finite, i.e. cutoff independent
lagrangian.
The cutoff appears in an explicit manner in the relativistic
path integral, (\ref{pathiml}), as in bare Quantum Field Theory.

There are two scale parameters in our path integrals, the cutoff, 
$c\Delta t$ and $\lambda_C>>c\Delta t$. We shall first consider the
asymptotically ultraviolet regime, for the length scales above the
cutoff but safely below the Compton wave length. In this regime
the mass can be neglected or treated as a weak perturbation 
and the propagation is practically governed by
the neutrino equation. In particular, it has been pointed out in ref. 
\cite{jacsch} and \cite{gjks} that the helicity flip occurs rarely
and the particle stays on the light cone in this regime. The insertion
of the covariant coordinate operator introduced in \cite{wigner} corresponds 
to the multiplication with the coordinate in the path integral. In this 
manner the acausal spread of the wave packet, (\ref{spread}), 
when the initial localization is comparable or better than 
$c\Delta t$ is the path integral analogue of the 
nonlocal feature of the covariant coordinate operator.
The $\hbar$ independence of the spread actually shows the kinematical origin.
The spread beyond the  Compton wavelength
looses its dependence on the cutoff and one recovers the usual
nonrelativistic, $\hbar$ dependent behavior. $\hbar$ enters via the
mass term and several helicity flips may occure simultaneously
in this regime. The Poisson distribution of the helicity flips
characterizing the relativistic domain gives rise the Wiener
process of the nonrelativistic propagation at the infrared side of 
the Compton wave length, \cite{jacsch}, \cite{gjks}.

The dependence of the average velocity on the observational time scale can be
traced down by the method of the renormalization group.
This amounts to perform the convolution (\ref{gblock})
repeatedly and to follow the average velocity as the time of
the observation, $t$, increases. The dependence obtained in this manner,
\be
\left({\Delta x\over t}\right)^2=O\left({1\over t}\right),\label{brown}
\ee
can be interpreted as a scaling relation.

An analogous scaling law results for the
free relativistic particles since they lagrangian is quadratic.
The massless propagator, (\ref{kernelft}), is nonvanishing for $t>>\Delta
t$ when
\be
\left({\Delta x\over t}\right)^2\approx c^2,\label{relscl}
\ee
since the eigenvalue of the Pauli matrices is $\pm1$. For $t\approx\Delta t$
one finds
\be
\left({\Delta x\over t}\right)^2\approx c^2{\Delta t\over t}.
\ee
Thus we recover the usual kinematics for the propagation of the massless
particles at time scales longer than the cutoff.

The asymptotic scaling agree for massive and massless particles
in the ultraviolet regime. Starting from the infrared end 
the usual nonrelativistic scaling law must somehow change into the 
relativistic one for a massive particle where the average velocity reaches
the speed of the light, $ct\approx\lambda_C$. To see this
on a qualitative level we shall evaluate the average velocity
by neglecting the influence of the end points of the trajectories
in the path integral,
\be
<\left({\Delta\s x\over ct}\right)^2>=\int d\Delta\s xG(\Delta\s x,0;t)
\left({\Delta\s x\over ct}\right)^2.\label{avvel}
\ee
We employ the semiclassical approximation to the integral in the propagator
(\ref{print}) and in (\ref{avvel}).
The extremum of the exponent in (\ref{print}) given by
\be
{|\s x|\over ct}={|\s p|\over mc}{1\over\sqrt{1+\left({\s p\over
mc}\right)^2}},
\ee
which shows that the contribution is concentrated in the causal region,
\be
{|\s x|\over ct}<1.\label{causal}
\ee
The value of the phase factor at the saddle point,
\be
\exp\left(i{ct\over\lambda_C}\sqrt{1-\left({\s x\over ct}\right)^2}\right),
\ee
is the integrand for the $\s x$ integration in the range (\ref{causal})
what covers the possible saddle points. The integral (\ref{avvel}) is
rewritten
by means of the dimensionless variable $\s z=\s x/ct$,
\bea
<\left({\Delta\s x\over ct}\right)^2>&=&
{\int d\s z\s z^2e^{i{ct\over\lambda_C}\sqrt{1-\s z^2}}\over
\int d\s ze^{i{ct\over\lambda_C}\sqrt{1-\s z^2}}}\nonu
&\approx&\cases{O(t^0)&for $tc<<\lambda_C$,\cr
{\hbar\over mc^2t}&for $tc>>\lambda_C$,\cr}
\eea
in leading order of $\hbar$, indicating a crossover at the
relativistic-nonrelativistic transition at $\Delta tc\approx\lambda_C$. In
fact, the
path integral displays different universal
behavior in these regimes: The trajectories of the path integral follow the
massless relativistic scaling laws for $ct<\lambda_C$ and appear to be
differentiable, $|{\Delta x\over t}|\approx c$. The usual nonrelativistic
scaling law
characteristic of the Brownian motion sets in for $ct>\lambda_C$ .

\section{Summary}
A path integral was presented in this paper which reproduces the solutions of
the Klein-Gordon and the Dirac equations. The nonrelativistic limit was
verified
for massive fermions. It was found that the trajectories of the path
integral follow the nonrelativistic scaling law characterizing the Brownian
motion when $c\Delta t>\lambda_C$. When the observational time, $\Delta t$, is
decreased down to the value $\Delta t\approx\lambda_C/c$ then the average velocity
reaches the order of magnitude of
$c$ and the relativistic effects become visible as a crossover. In the
relativistic regime, $c\Delta t<<\lambda_C$, the average velocity remains $c$.

The free particle corresponds to a quadratic lagrangian
$L=L_1+L_2$, $L_1=O(\Delta x)$, $L_2=O(\Delta^2x)$
where the linear terms were responsible for the time evolution. The
coefficient of
the quadratic term is linearly divergent, $L_2=O((\Delta t)^{-1})$,
and appeared as a formal device only. It is worthwhile
noting that the explicit appearance of the linear divergent coefficient,
$(\Delta t)^{-1}$, does not mean that this path integral is more
singular in the continuum limit that its nonrelativistic counterpart. Actually
the situation is just the contrary. Note that in both cases the phase angle
$\Delta tL=O((\Delta t)^0)$ for the typical trajectories. The diverging
coefficient in the relativistic lagrangian suppresses the velocity
fluctuations and renders the trajectories differentiable. In such a manner
the relativistic effects "regulate" the It\^o calculus \cite{ito} and
the classical calculus is recovered within the relativistic regime.

The spin-statistic theorem asserts the equivalence of the phase factor
appearing in front of the wave function under rotation by $2\pi$ and
exchange of two particles. In this manner one may speculate that the proper
treatment of the spin half aspect of the propagation which
includes the -1 factor for each rotation by $2\pi$ will
ultimately yield the correct antisymmetrized Green functions
in the second quantization without the use of formal Grassmann variables
in the path integral.

\end{document}